\documentclass[conference,a4paper]{IEEEtran}

\usepackage{cite}

\usepackage{latexsym}
\usepackage{graphics}
\usepackage{amssymb}
\usepackage{amsthm}
\usepackage{amsmath}

\usepackage{cite}
\usepackage{array}
\usepackage{mdwmath}
\usepackage{mdwtab}
\usepackage[tight,footnotesize]{subfigure}
\usepackage{graphicx}

\def\tr{\mathrm{tr}}
\def\trace{\mathrm{tr}}
\def\rank{\mathrm{rank}}

\def\diag{\mathrm{diag}}
\def\NBS{N_{\text{BS}}}
\def\NSCA{N_{\text{SCA}}}
\def\Si{j}
\def\vark{i}

\newcommand{\condSum}[3]{\overset{#3}{\underset{\underset{#2}{#1}}{\sum}}}

\newcommand{\fracSumtwo}[2]{\overset{#2}{\underset{#1}{\sum}}}
\newcommand{\vect}[1]{\mathbf{#1}}

\newcommand{\minimize}[1]{{\underset{{#1}}{\mathrm{minimize}}}}

\theoremstyle{plain}
\newtheorem{remark}{Remark}

\newtheorem{theorem}{Theorem}
\newtheorem{corollary}{Corollary}

\begin{document}

\title{Massive MIMO and Small Cells: Improving Energy Efficiency by Optimal Soft-Cell Coordination}

\IEEEoverridecommandlockouts

\author{\IEEEauthorblockN{Emil Bj{\"o}rnson\IEEEauthorrefmark{1}\IEEEauthorrefmark{2},
Marios Kountouris\IEEEauthorrefmark{3}, and
M{\'e}rouane Debbah\IEEEauthorrefmark{1} \thanks{E. Bj\"ornson is funded by the International Postdoc Grant 2012-228 from The Swedish Research Council. This research has been supported by the ERC Starting Grant 305123 MORE (Advanced Mathematical Tools for Complex Network Engineering).}}
\IEEEauthorblockA{\IEEEauthorrefmark{1}Alcatel-Lucent Chair on Flexible Radio, SUPELEC, Gif-sur-Yvette, France}
\IEEEauthorblockA{\IEEEauthorrefmark{3}Department of Telecommunications, SUPELEC, Gif-sur-Yvette, France}
\IEEEauthorblockA{\IEEEauthorrefmark{2}ACCESS Linnaeus Centre, Signal Processing Lab, KTH Royal Institute of Technology, Stockholm, Sweden\\
Emails: \{emil.bjornson, marios.kountouris, merouane.debbah\}@supelec.fr}}

\maketitle

\begin{abstract}
To improve the cellular energy efficiency, without sacrificing quality-of-service (QoS) at the users, the network topology must be densified to enable higher spatial reuse. We analyze a combination of two densification approaches, namely ``massive'' multiple-input multiple-output (MIMO) base stations and small-cell access points. If the latter are operator-deployed, a spatial soft-cell approach can be taken where the multiple transmitters serve the users by joint non-coherent multiflow beamforming. We minimize the total power consumption (both dynamic emitted power and static hardware power) while satisfying QoS constraints. This problem is proved to have a hidden convexity that enables efficient solution algorithms. Interestingly, the optimal solution promotes exclusive assignment of users to transmitters. Furthermore, we provide promising simulation results showing how the total power consumption can be greatly improved by combining massive MIMO and small cells; this is possible with both optimal and low-complexity beamforming.
\end{abstract}

\IEEEpeerreviewmaketitle

\section{Introduction}

The classical macro-cell network topology is well-suited for providing wide-area coverage, but cannot handle the rapidly increasing user numbers and QoS expectations that we see today---the energy efficiency would be very low. The road forward seems to be a densified topology that enables very high spatial reuse. Two main approaches are currently investigated: \emph{massive MIMO} \cite{Rusek2012a,Hoydis2013a} and \emph{small-cell networks} \cite{Parkvall2011a,Hoydis2011c}.

The first approach is to deploy large-scale antenna arrays at existing macro base stations (BSs) \cite{Rusek2012a}. This enables precise focusing of emitted energy on the intended users, resulting in a much higher energy efficiency. The channel acquisition is indispensable for massive MIMO, which requires the exploitation of channel reciprocity using time-division duplex (TDD). This mode makes the channel estimation accuracy limited by the number of users and not the number of BS antennas \cite{Rusek2012a}.

The second approach is to deploy an overlaid layer of small-cell access points (SCAs) to offload traffic from BSs, thus exploiting the fact that most data traffic is localized and requested by low-mobility users. This approach reduces the average distance between users and transmitters, which translates into lower propagation losses and higher energy efficiency \cite{Hoydis2011c}. This comes at the price of having a highly heterogeneous network topology where it is difficult to control and coordinate inter-user interference. To meet this challenge, industry \cite{Parkvall2011a} and academia \cite{Hoydis2011c} are shifting focus from user-deployed femtocells to operator-deployed SCAs. The latter can rely on reliable backhaul connectivity and joint control/coordination of BS and SCAs; the existence of SCAs can even be transparent to the users, as in the soft-cell approach proposed for LTE in \cite{Parkvall2011a}.

The total power consumption can be modeled with a static part that depends on the transceiver hardware and a dynamic part which is proportional to the emitted signal power \cite{Cui2005a,EARTH_D23,Ng2012a}. Massive MIMO and small-cell networks promise great improvements in the dynamic part, but require more hardware and will therefore increase the static part. In other words, dense network topologies must be properly deployed and optimized to actually improve the overall energy efficiency.

This paper analyzes the possible improvements in energy efficiency when the classical macro-cell topology is modified by employing massive MIMO at the BS and/or overlaying with SCAs. We assume perfect channel acquisition and a backhaul network that supports interference coordination; we thus consider an ultimate bound on what is practically achievable. The goal is to \emph{minimize the total power consumption while satisfying QoS constraints at the users and power constraints at the BS and SCAs}. We show that this optimization problem has a hidden convex structure that enables finding the optimal solution in polynomial time. The solution is proved to automatically/dynamically assign each user to the optimal transmitter (BS or SCA). A low-complexity algorithm based on classical regularized zero-forcing (RZF) beamforming is proposed and compared with the optimal solution. The potential merits of different densified topologies are analyzed by simulations.

\section{System Model}

\begin{figure}[t!]
\includegraphics[width=\columnwidth]{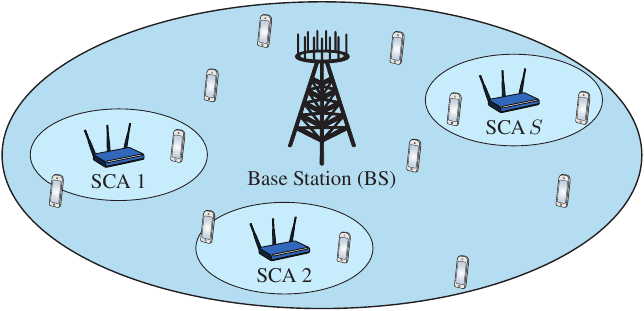} \vskip -3mm
\caption{Illustration of a downlink macro-cell overlaid with $S$ small cells. The BS has $\NBS$ antennas and the SCAs have $\NSCA$ antennas. The $K$ single-antenna users (e.g., smartphones) can be served (non-coherently) by any combination of transmitters, but the circles indicate typical coverage areas.}\label{figure_small_cell_overlay} \vskip -4mm
\end{figure}

We consider a single-cell downlink scenario where a macro BS equipped with $\NBS$ antennas should deliver information to $K$ single-antenna users. In addition, there are $S \geq 0$ SCAs that form an overlay layer and are arbitrarily deployed. The SCAs are equipped with $\NSCA$ antennas each, typically $1 \leq \NSCA \leq  4$, and characterized by strict power constraints that limit their coverage area (see below). In comparison, the BS has generous power constraints that can support high QoS targets in a large coverage area. The number of antennas, $\NBS$, is anything from 8 to several hundred---the latter means that $\NBS \gg K$ and is known as massive MIMO. This scenario is illustrated in Fig.~\ref{figure_small_cell_overlay}.

The channels to user $k$ are modeled as block fading. We consider a single flat-fading subcarrier where the channels are represented in the baseband by $\vect{h}_{k,0}^H \in \mathbb{C}^{1 \times \NBS}$ and $\vect{h}_{k,\Si}^H \in \mathbb{C}^{1 \times \NSCA}$ for the BS and $\Si$th SCA, respectively. These are assumed to be perfectly known at both sides of each channel; extensions with robustness to channel uncertainty can be obtained as in \cite{Bjornson2013d}. The received signal at user $k$ is
\begin{equation}
y_{k} = \vect{h}_{k,0}^H \vect{x}_0 + \sum_{\Si=1}^{S} \vect{h}_{k,\Si}^H \vect{x}_{\Si} + n_k
\end{equation}
\noindent where $\vect{x}_0,\vect{x}_{\Si}$ are the transmitted signals at the BS and $\Si$th SCA, respectively. The term $n_k \sim \mathcal{CN}(0,\sigma_k^2)$ is the circularly-symmetric complex Gaussian receiver noise with zero-mean and variance $\sigma_k^2$, measured in milliwatt (mW).

The BS and SCAs are connected to a backhaul network that enables joint spatial soft-cell resource allocation but only linear non-coherent transmissions; that is, each user can be served by multiple transmitters but the information symbols will be coded and emitted independently. We call it \emph{spatial multiflow transmission} \cite{Holma2012a} and it enables users barely covered by a SCA to receive extra signals from the BS or other SCAs.

The information symbols from the BS and the $\Si$th SCA to user $k$ are denoted $x_{k,0}$ and $x_{k,\Si}$, respectively, and originate from independent Gaussian codebooks with unit power (in mW); that is, $x_{k,\Si} \sim \mathcal{CN}(0,1)$ for $\Si=0,\ldots,S$. These symbols are multiplied with the beamforming vectors $\vect{w}_{k,0} \in \mathbb{C}^{\NBS \times 1}$ and $\vect{w}_{k,\Si} \in \mathbb{C}^{\NSCA \times 1}$ to obtain the transmitted signals
\begin{equation}
\vect{x}_{\Si} = \sum_{k=1}^{K} \vect{w}_{k,\Si} x_{k,\Si},\quad \Si=0,\ldots,S.
\end{equation}
The beamforming vectors are the optimization variables in this paper. Note that $\vect{w}_{k,\Si}\neq \vect{0}$ only for transmitters $j$ that serve user $k$. This transmitter assignment is obtained automatically and optimally from the optimization problem solved herein.

\subsection{Problem Formulation}

This paper considers minimization of the total power consumption while satisfying QoS constraints for each user. We will define both concepts before formulating the problem.

The QoS constraints specify the information rate [bits/s/Hz] that each user should achieve in parallel. These are defined as $\log_2(1+\textrm{SINR}_k) \geq \gamma_k$, where $\gamma_k$ is the fixed QoS target and
\begin{equation} \label{eq_achievable_rate}
\textrm{SINR}_k =\frac{ |\vect{h}_{k,0}^H \vect{w}_{k,0}|^2 + \fracSumtwo{\Si=1}{S} |\vect{h}_{k,\Si}^H \vect{w}_{k,\Si}|^2 }{ \condSum{\vark=1}{\vark \neq k}{K} \Big( |\vect{h}_{k,0}^H \vect{w}_{\vark,0}|^2 + \fracSumtwo{\Si=1}{S} |\vect{h}_{k,\Si}^H \vect{w}_{\vark,\Si}|^2 \Big)  + \sigma_k^2 }
\end{equation}
is the aggregate signal-to-interference-and-noise ratio (SINR) of the $k$th user. The information rate $\log_2(1+\textrm{SINR}_k)$ is achieved by applying successive interference cancellation on the own information symbols and treating co-user symbols as noise. Observe that this rate is obtained without any phase-synchronization between transmitters, contrary to coherent joint transmission that requires very tight synchronization \cite{Bjornson2011a}.

The power consumption (per subcarrier) can be modeled as $P_{\text{dynamic}}+P_{\text{static}}$ \cite{Cui2005a,EARTH_D23,Ng2012a} with the dynamic and static terms
\begin{align}
P_{\text{dynamic}} &= \rho_0 \sum_{k=1}^{K} \| \vect{w}_{k,0} \|^2 + \sum_{\Si=1}^{S} \rho_{\Si} \sum_{k=1}^{K} \| \vect{w}_{k,\Si} \|^2, \\
P_{\text{static}} &= \frac{\eta_0}{C}  \NBS + \sum_{\Si=1}^{S} \frac{\eta_{\Si}}{C} \NSCA,
\end{align}
respectively. The dynamic term is the aggregation of the emitted powers, $\sum_{k=1}^{K} \| \vect{w}_{k,\Si} \|^2$, each multiplied with a constant $\rho_{\Si} \geq 1$ accounting for the inefficiency of the power amplifier at this transmitter. The static term, $P_{\text{static}}$, is proportional to the number of antennas and $\eta_{\Si} \geq 0$ models the power dissipation in the circuits of each antenna (e.g., in filters, mixers, converters, and baseband processing). $P_{\text{static}}$ is normalized with the total number of subcarriers $C \geq 1$. Representative numbers on these parameters are given in Table \ref{table_parameters_hardware}, \cite{EARTH_D23}, and \cite{Kumar2011a}

Each BS and SCA is prone to $L_{\Si}$ power constraints
\begin{equation} \label{eq_power_constaints}
\sum_{k=1}^{K} \vect{w}_{k,\Si}^H \vect{Q}_{\Si,\ell} \vect{w}_{k,\Si} \leq q_{\Si,\ell}, \quad \ell=1,\ldots,L_{\Si}.
\end{equation}
The weighting matrices $\vect{Q}_{0,\ell} \in \mathbb{C}^{\NBS \times \NBS},\vect{Q}_{\Si,\ell} \in \mathbb{C}^{\NSCA \times \NSCA}$ for $\Si=1,\ldots,S$, are positive semi-definite. The corresponding limits are $q_{\Si,\ell}\geq 0$. The parameters $\vect{Q}_{\Si,\ell},q_{\Si,\ell}$ are fixed and can describe any combination of per-antenna, per-array, and soft-shaping constraints \cite{Bjornson2011a}. We typically have $q_{0,\ell} \gg q_{\Si,\ell}$ for $1\leq \Si \leq S$, because the BS provides coverage. Our numerical evaluation considers per-antenna constraints of $q_{\Si}$ [mW] at the $\Si$th transmitter, given by $L_{0}=\NBS$, $L_{\Si}=\NSCA$, $q_{\Si,\ell} = q_{\Si} \,\, \forall \ell$, and $\vect{Q}_{\Si,\ell}$ with one at $\ell$th diagonal element and zero elsewhere.

We are now ready to formulate our optimization problem. We want to minimize the total power consumption while satisfying the QoS constraints and the power constraints, thus \vskip-3mm
\begin{equation} \label{eq_global_problem}
\begin{split}
\minimize{\vect{w}_{k,\Si} \, \forall k,\Si} \,\,\,& \,\, P_{\text{dynamic}} + P_{\text{static}} \\
\mathrm{subject \,\,to} \,\, & \,\, \log_2(1+\textrm{SINR}_k) \geq \gamma_k \quad \forall k, \\
& \,\, \sum_{k=1}^{K} \vect{w}_{k,\Si}^H \vect{Q}_{\Si,\ell} \vect{w}_{k,\Si} \leq q_{\Si,\ell} \quad \forall \Si,\ell.
\end{split}
\end{equation}

In the next section, we will prove that \eqref{eq_global_problem} can be reformulated as a convex optimization problem and thus is solvable in polynomial time using standard algorithms. Moreover, the optimal power-minimizing solution is self-organizing in the sense that only one or a few transmitters will serve each user.

\begin{remark}
The static part, $P_{\text{static}}$, of the power consumption depends on the number of SCAs and antennas. From an energy efficiency perspective, it therefore makes sense to put inactive SCAs and antenna elements into sleep mode. On the other hand, such adaptive sleep mode techniques make the sensing of user mobility and new users complicated. There is also a non-negligible transient behavior when switching from sleep mode to active mode \cite{Cui2005a}. Since these problems are outside the scope of this paper, we will instead compare setups with different values on $\NBS$, $\NSCA$, and $S$ by using simulations.
\end{remark}

\section{Algorithms for Non-Coherent Coordination}

\label{section_algorithms}

This section derives algorithms for solving the optimization problem \eqref{eq_global_problem}. The QoS constraints in \eqref{eq_global_problem} are complicated functions of the beamforming vectors, making the problem non-convex in its original formulation. However, we will prove that it has an underlying convex structure that can be extracted using semi-definite relaxation. We generalize the original approach in \cite{Bengtsson2001a} to spatial multiflow transmission.

To achieve a convex reformulation of \eqref{eq_global_problem}, we use the notation $\vect{W}_{k,\Si}=\vect{w}_{k,\Si} \vect{w}_{k,\Si}^H \,\, \forall k,\Si$. This matrix should be positive semi-definite, denoted as $\vect{W}_{k,\Si} \succeq \vect{0}$, and have $\rank(\vect{W}_{k,\Si}) \leq 1$. Note that the rank can be zero, which implies that $\vect{W}_{k,\Si}=\vect{0}$. By including the BS and SCAs in the same sum expressions, we can rewrite \eqref{eq_global_problem} compactly as \vskip-4mm
\begin{align} \label{eq_global_problem_SDP}
\minimize{\vect{W}_{k,\Si} \succeq \vect{0} \,\, \forall k,\Si} \,\,\,& \,\, \sum_{\Si=0}^{S} \rho_{\Si} \sum_{k=1}^{K} \tr( \vect{W}_{k,\Si} ) + P_{\text{static}} \\ \notag
\mathrm{subject \,\,to} \,\,\,\, & \,\, \rank(\vect{W}_{k,\Si}) \leq 1 \quad \forall k,\Si, \\[-1mm] \notag
 \fracSumtwo{\Si=0}{S} \vect{h}_{k,\Si}^H & \bigg( \Big(1\!+\!\frac{1}{\tilde{\gamma}_k}\Big) \vect{W}_{k,\Si} - \sum_{\vark=1}^{K}  \vect{W}_{\vark,\Si} \bigg) \vect{h}_{k,\Si} \geq \sigma_k^2 \quad \forall k, \\[-2mm] \notag
& \,\, \sum_{k=1}^{K} \tr( \vect{Q}_{\Si,\ell} \vect{W}_{k,\Si}) \leq q_{\Si,\ell} \quad \forall \Si,\ell,
\end{align} \vskip-1mm
\noindent where the QoS targets have been transformed into SINR targets of $\tilde{\gamma}_k=2^{\gamma_k}-1 \,\, \forall k$. The problem \eqref{eq_global_problem_SDP} is convex except for the rank constraints, but we will now prove that these constraints can be relaxed without losing optimality.

\begin{theorem} \label{theorem_sdp_is_tight}
Consider the semi-definite relaxation of \eqref{eq_global_problem_SDP} where the rank constraints $\rank(\vect{W}_{k,\Si}) \leq 1$ are removed. This becomes a convex semi-definite optimization problem. Furthermore, it will always have an optimal solution $\{\vect{W}_{k,\Si}^* \,\, \forall k,\Si\}$ where all matrices satisfy $\rank(\vect{W}^*_{k,\Si}) \leq 1$.
\end{theorem}
\begin{IEEEproof}
The proof is given in the appendix.
\end{IEEEproof}

This theorem shows that the original problem \eqref{eq_global_problem} can be solved as a convex optimization problem. This means that the optimal solution is guaranteed in polynomial time \cite{Boyd2004a}; for example, using the interior-point toolbox \texttt{SeDuMi} \cite{SEDUMI}.

Further structure of the optimal solution can be obtained.

\begin{corollary} \label{cor_user_allocation}
Consider the optimal solution $\{\vect{W}_{k,\Si}^* \,\, \forall k,\Si\}$ to \eqref{eq_global_problem_SDP}. For each user $k$ there are three possibilities:
\begin{enumerate}
\item It is only served by the BS (i.e., $\vect{W}_{k,\Si}^*= \vect{0}$,  $1 \leq \Si \leq S$);
\item It is only served by the $\Si$th SCA (i.e., $\vect{W}_{k,0}^*= \vect{0}$  and $\vect{W}_{k,i}^*= \vect{0}$ for $i \neq \Si$);
\item It is served by a combination of BS and SCAs, whereof at least one transmitter $j$ has an active power constraint $\ell$ (i.e., $\sum_{k=1}^{K} \tr( \vect{Q}_{\Si,\ell} \vect{W}_{k,\Si}^*) = q_{\Si,\ell}$).
\end{enumerate}
\end{corollary}
\begin{IEEEproof}
The proof is given in the appendix.
\end{IEEEproof}

This corollary shows that although users can be served by multiflow transmission, it is usually optimal to assign one transmitter per user. Users that are close to a SCA are served exclusively by it, while most other users are served by the BS. There are transition areas around each SCA where multiflow transmission is utilized since the SCA is unable to fully support the QoS targets. Corollary \ref{cor_user_allocation} is a positive result since a reduced transmission/reception complexity is often optimal.

If the power constraints are removed, then the transition areas disappear. We refer to \cite{Bengtsson2001b} for prior work on dynamic transmitter assignment by means of convex optimization.

\subsection{Low-Complexity Algorithm}
\label{subsection_multiflowRZF}

The optimal beamforming for spatial soft-cell coordination can be computed in polynomial time using Theorem \ref{theorem_sdp_is_tight}. This complexity is relatively modest, but the algorithm becomes infeasible for real-time implementation when $\NBS$ and $S$ grow large. In addition, Theorem \ref{theorem_sdp_is_tight} provides a centralized algorithm that requires all channel knowledge to be gathered at the BS. Distributed algorithms can certainly be obtained using primal/dual decomposition techniques \cite{Bjornson2013d}, but these require iterative backhaul signaling of coupling variables---thus they are also infeasible for real-time implementations.

Theorem \ref{theorem_sdp_is_tight} should be seen as the ultimate benchmark when evaluating low-complexity algorithms for non-coherent coordination. To demonstrate the usefulness, we propose the low-complexity non-iterative \textbf{Multiflow-RZF beamforming}:

\begin{enumerate}
    \item Each transmitter $j=0,\ldots,S$ computes
    \begin{align*}
   \!\!\!\!   \vect{u}_{k,\Si} &= \frac{\big( \sum_{\vark=1}^{K} \frac{1}{\sigma_{\vark}^2} \vect{h}_{\vark,\Si} \vect{h}_{\vark,\Si}^H + \frac{K}{\tilde{\gamma}_k q_{\Si}} \vect{I} \big)^{-1} \vect{h}_{k,\Si}}{\big\| \big(\sum_{\vark=1}^{K} \frac{1}{\sigma_{\vark}^2} \vect{h}_{\vark,\Si} \vect{h}_{\vark,\Si}^H + \frac{K}{\tilde{\gamma}_k q_{\Si}} \vect{I} \big)^{-1} \vect{h}_{k,\Si} \big\|} \,\,\,  \forall k, \\
    \!\!\!\! g_{\vark,k,\Si} &= |\vect{h}_{\vark,\Si}^H \vect{u}_{k,\Si}|^2 \,\,\, \forall \vark,k, \quad Q_{\Si,\ell,k} =\vect{u}_{k,\Si}^H \vect{Q}_{\Si,\ell} \vect{u}_{k,\Si} \,\,\, \forall \ell,k.
    \end{align*}

    \item The $j$th SCA sends the scalars $g_{\vark,k,\Si},Q_{\Si,\ell,k} \, \forall k,\vark,\ell$ to the BS. The BS solves the convex optimization problem
\begin{align} \label{eq_power_alloc_problem}
\minimize{p_{k,\Si} \geq 0 \,\, \forall k,\Si} \,\,\,& \,\, \sum_{\Si=0}^{S} \rho_{\Si} \sum_{k=1}^{K} p_{k,\Si} + P_{\text{static}} \\  \notag
\mathrm{subject \,\,to} \,\,\, & \,\, \sum_{k=1}^{K} Q_{\Si,\ell,k} p_{k,\Si} \leq q_{\Si,\ell} \quad \forall \Si,\ell, \\ \notag
 \fracSumtwo{\Si=0}{S} \, p_{k,\Si} g_{k,k,\Si} &\Big(1\!+\!\frac{1}{\tilde{\gamma}_k}\Big) - \sum_{\vark=1}^{K}  p_{\vark,\Si}
 g_{k,\vark,\Si} \geq \sigma_k^2 \quad \forall k.
\end{align}

    \item The power allocation $p^*_{k,\Si} \,\, \forall k$ that solves \eqref{eq_power_alloc_problem} is sent to the $j$th SCA, which computes $\vect{w}_{k,\Si} = \sqrt{p^*_{k,\Si}} \vect{u}_{k,\Si} \,\, \forall k$.

\end{enumerate}

This algorithm applies the heuristic RZF beamforming (see e.g.,~\cite{Hoydis2013a}) to transform \eqref{eq_global_problem} into the power allocation problem \eqref{eq_power_alloc_problem}, which has the same low complexity irrespectively of the number of antennas. The algorithm is non-iterative, but some scalar parameters are exchanged between the BS and SCAs to enable coordination. In practice, only users in the vicinity of an SCA are affected by it, thus only a few parameters are exchanged per SCA while all other parameters are set to zero.

\section{Numerical Evaluations}
\label{section_numerical_illustrations}

\begin{figure}[t!]
\includegraphics[width=\columnwidth]{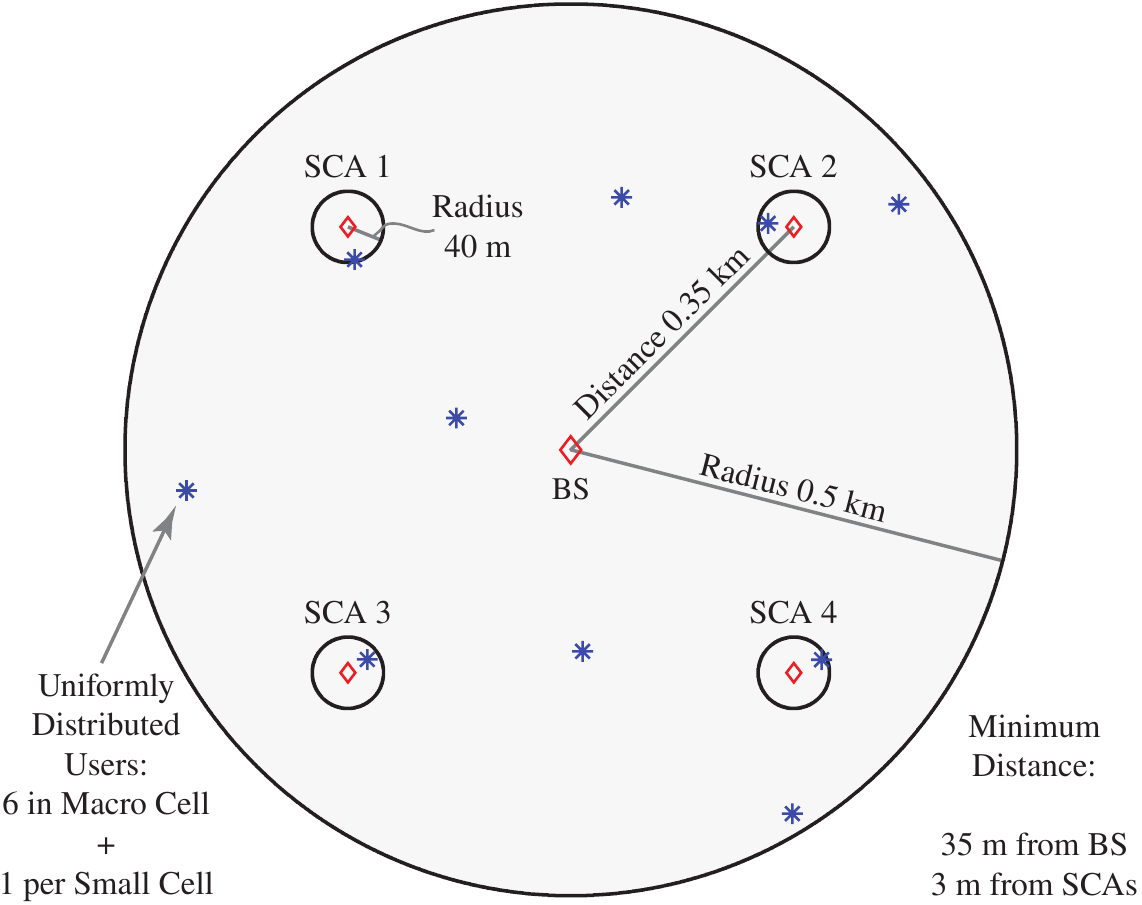} \vskip -2mm
\caption{The single-cell scenario analyzed in Section \ref{section_numerical_illustrations}. The BS and SCAs are fixed, while the 10 users are randomly distributed as described above.}\label{figure_simulation_scenario}
\end{figure}

This section illustrates the analytic results and algorithms of this paper in the scenario depicted in Fig.~\ref{figure_simulation_scenario}. This figure shows a circular macro cell overlaid by 4 small cells. There are 10 active users in the macro cell, whereof 6 users are uniformly distributed in the whole cell and each SCA has one user uniformly distributed within 40 meters. We evaluate the average performance over user locations and channel realizations. Table \ref{table_parameters_hardware} shows the hardware parameters that characterize the power consumption and is based on \cite[Table 7]{EARTH_D23} and \cite{Kumar2011a}.

\begin{table}[!t]
\renewcommand{\arraystretch}{1.3}
\caption{Hardware Parameters in the Numerical Evaluation}
\label{table_parameters_hardware} \vskip-2mm
\centering
\begin{tabular}{|c|c|}
\hline
\bfseries Parameters & \bfseries Values\\
\hline

Efficiency of power amplifiers & $\frac{1}{\rho_0}=0.388, \frac{1}{\rho_{\Si}}=0.052$ $\forall \Si$\\

Circuit power per antenna & $\eta_0 = 189$ mW, $\eta_{\Si} = 5.6$ mW $\forall \Si$\\

Per-antenna constraints & $q_{0,\ell} = 66$, $q_{\Si,\ell} = 0.08$ mW $\forall \Si,\ell$\\

\hline
\end{tabular} \vskip-3mm
\end{table}

The channels are modeled similarly to Case 1 for Heterogeneous deployments in the 3GPP LTE standard \cite{LTE2010b}, but the small-scale fading is modified to reflect recent works on massive MIMO. We assume Rayleigh small-scale fading: $\vect{h}_{k,\Si} \sim \mathcal{CN}(\vect{0},\vect{R}_{k,\Si})$. The correlation matrix is spatially uncorrelated, $\vect{R}_{k,\Si} \propto \vect{I}$, between the $\Si$th SCA and each user $k$. The correlation matrix between the BS and each user is modeled according to the physical channel model in \cite[Eq.~(34)]{Hoydis2013a}, where the main characteristics are antenna correlation and reduced-rank channels. Note that the propagation loss is different for BS and SCAs; see Table \ref{table_channel_parameters} for all channel model parameters.

\begin{table}[!t]
\renewcommand{\arraystretch}{1.3}
\caption{Channel Parameters in the Numerical Evaluation}
\label{table_channel_parameters} \vskip-2mm
\centering
\begin{tabular}{|c|c|}
\hline
\bfseries Parameters & \bfseries Values\\
\hline

Macro cell radius & 0.5 km \\

Carrier frequency / Number of subcarriers & $F=2$ GHz / $C=600$ \\

Total bandwidth / Subcarrier bandwidth & 10 MHz  / 15 kHz \\

Small-scale fading distribution & $\vect{h}_{k,\Si} \sim \mathcal{CN}(\vect{0},\vect{R}_{k,\Si})$ \\

Standard deviation of log-normal shadowing & 7 dB \\

Path and penetration loss at distance $d$ (km) & \!\!$148.1 \!+\! 37.6 \log_{10}(d)$ dB\!\!\\

Special case: Within 40 m from SCA & \!$127 \!+\! 30 \log_{10}(d)$ dB\!\\

Noise variance $\sigma_k^2$ (5 dB noise figure) & $-127$ dBm \\

\hline
\end{tabular} \vskip-3mm
\end{table}

\begin{figure}[t!]
\includegraphics[width=\columnwidth]{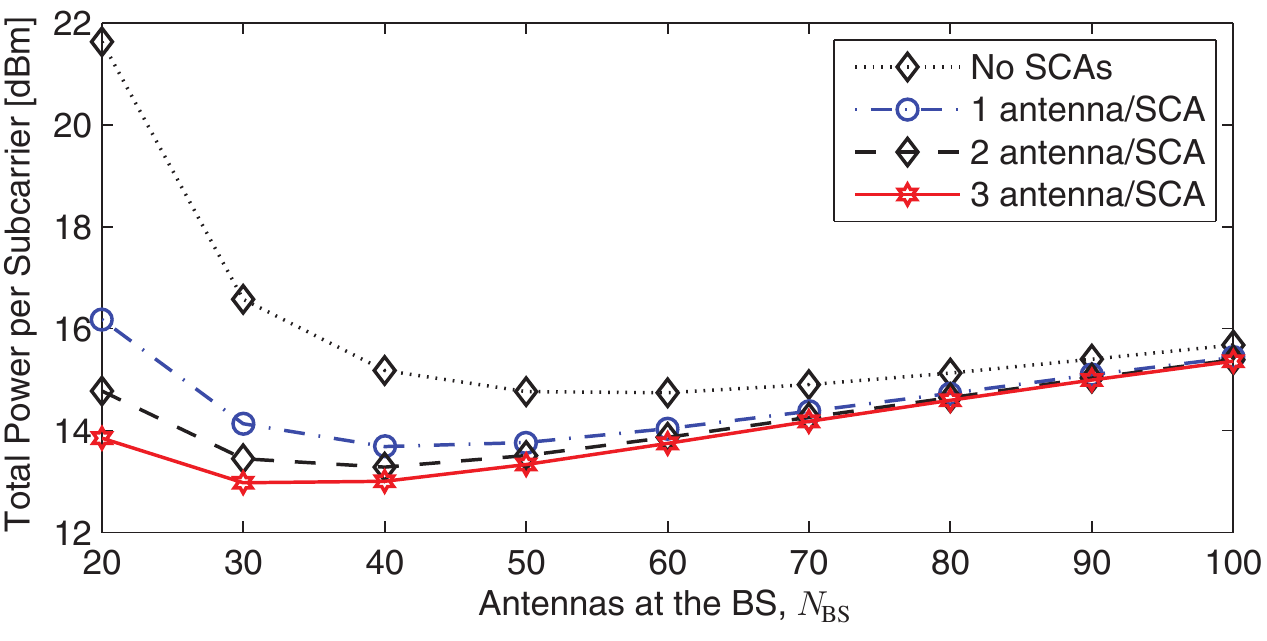} \vskip -3mm
\caption{Average total power consumption in the scenario of Fig.~\ref{figure_simulation_scenario}. We consider different $\NBS$ and $\NSCA$, while the QoS constraints are 2 bits/s/Hz.}\label{figure_totalpower} \vskip -5mm
\end{figure}

We first analyze the impact of having different number of antennas at the BS and SCAs: $ \NBS \in \{20,30,\ldots,100\}$, $\NSCA \in \{0,1,2,3\}$. Fig.~\ref{figure_totalpower} shows the average total power consumption (per subcarrier) in a scenario where the 10 users have QoS constraints of 2 bits/s/Hz. The optimal spatial multiflow transmission is obtained using Theorem \ref{theorem_sdp_is_tight} and the convex optimization problems were solved by the algorithmic toolbox \texttt{SeDuMi} \cite{SEDUMI}, using the modeling language \texttt{CVX} \cite{cvx2012}.

Fig.~\ref{figure_totalpower} demonstrates that adding more hardware can substantially decrease the total power consumption $P_{\text{dynamic}} + P_{\text{static}}$. This means that the decrease in the dynamic part, $P_{\text{dynamic}}$, due to better energy-focusing and less propagation losses clearly outweigh the increase in the static part, $P_{\text{static}}$, from the extra circuitry. Massive MIMO brings large energy efficiency improvements by itself, but the same power consumption can be achieved with half the number of BS antennas (or less) by deploying a few single-antenna SCAs in areas with active users. Further improvements in energy efficiency are achieved by having multi-antenna SCAs; a network topology that combines massive MIMO and small cells is desirable to achieve high energy efficiency with little additional hardware. However, there are saturation points where extra hardware will not decrease the total power anymore.
Note that the power is shown in dBm, thus there are 10-fold improvements in Fig.~\ref{figure_totalpower}.

Although the system allows for multiflow transmission, the simulation shows only a 0--3\% probability of serving a user by multiple transmitters. This is in line with Corollary \ref{cor_user_allocation}. The main impact of increasing $\NSCA$ is that each SCA is likely to being allocated more than one user to serve exclusively; the probability is 20--45\% for $\NSCA\!=\!3$ but decreases with $\NBS$.

\begin{figure}[t!]
\includegraphics[width=\columnwidth]{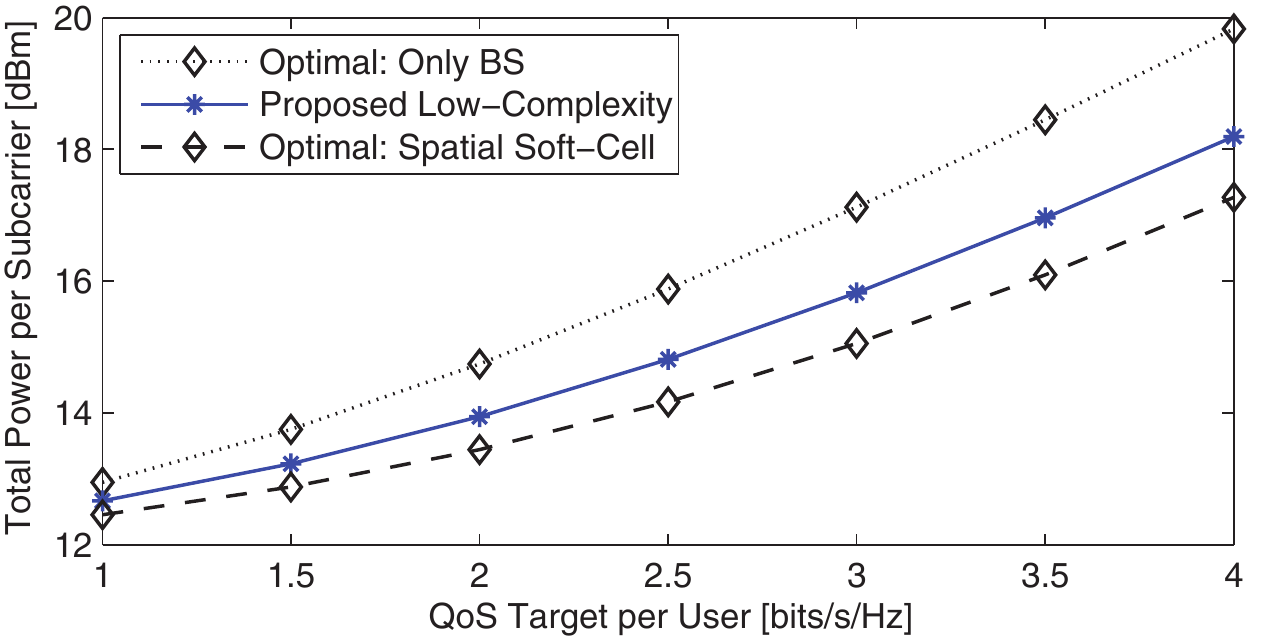} \vskip -3mm
\caption{Average total power consumption in the scenario of Fig.~\ref{figure_simulation_scenario} with $\NBS=50$ and $\NSCA=2$. We consider different QoS constraints and beamforming.}\label{figure_lowcomplexity} \vskip -4mm
\end{figure}

Next, Fig.~\ref{figure_lowcomplexity} considers $\NBS=50$ and $\NSCA=2$ for different QoS constraints. Three beamforming algorithms are compared: 1) Optimal beamforming using only the BS; 2) Multiflow-RZF proposed in Section \ref{subsection_multiflowRZF}; and 3) Optimal spatial soft-cell coordination from Theorem \ref{theorem_sdp_is_tight}.
As in the previous figure, we observe great improvements in energy efficiency by offloading users to the SCAs. The proposed Multiflow-RZF beamforming gives promising results for practical applications, because a majority of the energy efficiency improvements is achievable by judicious low-complexity beamforming techniques.

\section{Conclusion}

The energy efficiency of cellular networks can be improved by employing massive MIMO at the BSs or overlaying current infrastructure by a layer of SCAs. This paper analyzed a combination of these concepts based on soft-cell coordination, where each user can be served by non-coherent beamforming from multiple transmitters. We proved that the \emph{power-minimizing spatial multiflow transmission under QoS constraints} is achieved by solving a convex optimization problem. The optimal solution dynamically assigns users to the optimal transmitters, which usually is only the BS or one of the SCAs.

The analysis considered both the dynamic emitted power and static hardware consumption. We provide promising results showing that the \emph{total power consumption} can be greatly improved by combining massive MIMO and small cells. Most of the benefits are also achievable by low-complexity beamforming, such as the proposed Multiflow-RZF beamforming.

\appendix

\label{section_appendix}

\textbf{Proof of Theorem \ref{theorem_sdp_is_tight}.} The relaxed problem is a semi-definite optimization problem on standard form \cite{Boyd2004a}. As shown in \cite[Example 1]{Bjornson2011a}, there might exist high-rank solutions. However, there always exist a solution with $\rank(\vect{W}^*_{k,\Si}) \leq 1 \,\, \forall k,\Si$. To prove this, suppose there exist an optimal solution $\{\vect{W}_{k,\Si}^{**} \,\, \forall k,\Si\}$ with $\rank(\vect{W}_{k,\Si}^{**})>1$ for some $k,\Si$. We can replace $\vect{W}_{k,\Si}^{**}$ by any $\vect{V} \succeq \vect{0}$ that maximizes $\vect{h}_{k,\Si}^H  \vect{V}\, \vect{h}_{k,\Si}$ subject to $\trace(\vect{V}) \leq \trace(\vect{W}^{**}_{k,\Si})$, $\tr( \vect{Q}_{\Si,\ell} \vect{V}) \leq \tr( \vect{Q}_{\Si,\ell} \vect{W}^{**}_{k,\Si}) \,\, \forall \ell$ (i.e., not using more power than $\vect{W}_{k,\Si}^{**}$) and $\vect{h}_{\vark,\Si}^H \vect{V} \,\vect{h}_{\vark,\Si} \leq \vect{h}_{\vark,\Si}^H \vect{W}^{**}_{k,\Si} \vect{h}_{\vark,\Si} \,\, \forall \vark \neq k$ (i.e., not causing more interference than $\vect{W}_{k,\Si}^{**}$). One solution is $\vect{V}=\vect{W}_{k,\Si}^{**}$, but \cite[Lemma 3]{Bjornson2011a} shows that problems of this form always have rank-one solutions.

\textbf{Proof of Corollary \ref{cor_user_allocation}.} For convenience, let $\vect{A}_k = \frac{1}{\sigma_k^2} \diag( \frac{1}{\rho_0}\vect{h}_{k,0} \vect{h}_{k,0}^H, \ldots, \frac{1}{\rho_S} \vect{h}_{k,S} \vect{h}_{k,S})$ be a block-diagonal matrix and $\vect{w}_k = [ \sqrt{\rho_0} \vect{w}_{k,0}^T \, \ldots \, \sqrt{\rho_S} \vect{w}_{k,S}^T]^T$ be the aggregate beamforming vectors. Furthermore, let $\tilde{\vect{Q}}_{\Si,\ell} $ be the block-diagonal matrix that makes $\vect{w}_k^H \tilde{\vect{Q}}_{\Si,\ell} \vect{w}_k = \vect{w}_{k,\Si}^H \vect{Q}_{\Si,\ell} \vect{w}_{k,\Si}$.

Suppose $\vect{w}_k^* = \sqrt{p_k} \vect{u}_k$ is the optimal solution to \eqref{eq_global_problem}, where $\vect{u}_k$ is unit-norm.
By the uplink-downlink duality in \cite[Lemma 4]{Bjornson2011a}, we have \vskip-4mm
\begin{equation} \label{eq_uplinkdownlink_expression}
\tilde{\gamma}_k \!=\!\frac{ p_k \vect{u}_k^H \vect{A}_k \vect{u}_k}{\sum_{\vark \neq k} p_{\vark} \vect{u}_{\vark}^H \vect{A}_k \vect{u}_{\vark} \!+\! 1}
\!=\!\frac{ \lambda_k \vect{u}_k^H \vect{A}_k \vect{u}_k}{\vect{u}_{k}^H \vect{B}_k \vect{u}_{k}}
\end{equation}
where $\vect{B}_k=\big( \sum_{\vark \neq k} \lambda_{\vark} \vect{A}_{\vark}  \!+\! \sum_{\Si,\ell} \mu_{\Si,\ell} \tilde{\vect{Q}}_{\Si,\ell} \!+\! \vect{I} \big)$ and $\lambda_k,\mu_{\Si,\ell}$ are the optimal Lagrange multipliers for the QoS and power constraints, respectively.
The last expression in \eqref{eq_uplinkdownlink_expression}, the uplink SINR, takes its largest value when $\vect{u}_k$ is the dominating eigenvector of $\vect{B}_k^{-1/2} \vect{A}_k \vect{B}_k^{-1/2}$. Since $\vect{B}_k$ and $\vect{A}_k$ are block-diagonal, the dominating eigenvalue originates from one of the blocks and the corresponding eigenvector is only non-zero for this block. As each block corresponds to either the BS or one of the SCAs, this means that we ideally should serve user $k$ by only one transmitter.
The only reason to have another $\vect{u}_k$ is when there is multiplicity in the dominating eigenvalue and none of the single-transmitter solutions are supported by the power constraints; that is, when at least one power constraint is active. This proves the three cases stated in the corollary.

\bibliographystyle{IEEEtran}
\bibliography{IEEEabrv,refs}

\end{document}